\documentclass[a4paper]{revtex4}
\usepackage{graphicx}
\usepackage{fancyhdr}
\usepackage{amsmath}
\usepackage{color}

\begin{document}

\title{Leptogenesis during Axion Relaxation after Inflation}

%

\author{Kai Schmitz}
\affiliation{Kavli IPMU (WPI), University of Tokyo, Kashiwa, 277-8583, Japan}

\begin{abstract}
In this talk, I present a novel and minimal alternative to thermal leptogenesis,
which builds upon the assumption that the electroweak
gauge bosons are coupled to an axion-like scalar field, as it is, for instance,
the case in certain string compactifications.
The motion of this axion-like field after the end of inflation generates an effective chemical
potential for leptons and antileptons, which, in the presence of lepton number-violating
scatterings mediated by heavy Majorana neutrinos, provides an opportunity for
baryogenesis via leptogenesis.
In contrast to thermal leptogenesis, the final baryon asymmetry turns out to be insensitive
to the masses and $CP$-violating phases in the heavy neutrino sector.
Moreover, the proposed scenario requires a reheating temperature of at least
$\mathcal{O}(10^{12})\,\textrm{GeV}$ and it is, in particular,
consistent with heavy neutrino masses close the scale of grand unification.
This talk was given in February 2015 at \textit{HPNP 2015} at Toyama University
and is based on recent work (arXiv:1412.2043 [hep-ph]) in collaboration with
A.~Kusenko and T.~T.~Yanagida.
\end{abstract}

\maketitle

\thispagestyle{fancy}


\section{Introduction}

The standard model (SM) fails to provide an explanation for the baryon-to-photon
ratio in the present universe, $\eta_B^{\rm obs} \simeq 6 \times 10^{-10}$~\cite{Ade:2013zuv},
which serves as a major indication for new physics.
Consequently, some new dynamical mechanism must be responsible for baryogenesis, i.e., 
the generation of a primordial baryon-antibaryon asymmetry in the early universe~\cite{Dolgov:1997qr}.
Most mechanisms proposed in the literature are devised so as to satisfy the three
famous Sakharov conditions~\cite{Sakharov:1967dj}:
(i) violation of baryon number $B$ (or lepton number $L$ in the case of leptogenesis),
(ii) violation of $C$ as well as of $CP$ invariance, and
(iii) departure from thermal equilibrium.
As it turns out, it is, however, not mandatory to fulfill these three conditions
in order to successfully generate a baryon asymmetry.
The point is that Sakharov's conditions are based on the assumption of $CPT$ invariance,
which means that one is actually able to circumvent them in case $CPT$ is spontaneously broken.
This idea has been pioneered by Cohen and Kaplan in their scenario of \textit{spontaneous
baryogenesis}~\cite{Cohen:1987vi}, where the baryon asymmetry is generated in thermal equilibrium; and
since then, it has been studied and expanded upon by many
authors~\cite{Dine:1990fj,Kuzmin:1992up,Dolgov:1994zq,Chiba:2003vp}.
For example, Kusenko et al.\ have recently shown how the $CPT$ violation during the phase
of SM Higgs relaxation after the end of inflation can be used for the realization of baryogenesis via
leptogenesis~\cite{Kusenko:2014lra}.
In this talk, I will draw upon this earlier work and demonstrate that it can be easily generalized to
the case of generic axion-like scalar fields relaxing from large initial field values in the course
of reheating; further details pertaining to our analysis can be found in our
recent paper~\cite{Kusenko:2014uta} as well as in another forthcoming publication.


In an expanding universe at nonzero temperature, $CPT$ invariance can be easily broken
spontaneously by introducing a pseudoscalar field, $a(t,\vec{x})$, which
couples derivatively to the fermion current $j^\mu$ in the Lagrangian,
\begin{align}
\mathcal{L} \supset \frac{1}{f_a} \, \partial_\mu a \, j^\mu \,, \quad
j^\mu = \sum_f \bar{\psi}_f \gamma^\mu \psi_f \,,
\label{eq:derivative}
\end{align}
with $f_a$ being some cut-off scale.
Imposing spatial homogeneity, $a = a(t)$, and assuming that the classical
background is given cause to evolve with nonzero velocity, $\dot{a} \neq 0$,
(which is readily done in the early universe, as we will review shortly)
this coupling turns into an effective chemical potential
$\mu_{\rm eff}$ for the fermion number,
\begin{align}
\mathcal{L} \supset \frac{1}{f_a} \, \dot{a} \, j^0 = \mu_{\rm eff}\,  n_F \,, \quad
\mu_{\rm eff} = \frac{\dot{a}}{f_a} \,, \quad j^0 \equiv n_F = n_f - n_{\bar{f}}\,,
\label{eq:mueff}
\end{align}
which shifts the energy levels of fermions $f$ and antifermions $\bar{f}$ w.r.t.\ each other.
In thermal equilibrium, the minimum of the free energy is therefore obtained for a nonzero
fermion-antifermion asymmetry $n_F$, 
\begin{align}
n_{f,\bar{f}}^{\rm eq} \sim T^3 \left(1 \pm \frac{\mu_{\rm eff}}{T}\right) \,, \quad
n_F^{\rm eq} = n_f^{\rm eq} - n_{\bar{f}}^{\rm eq} \sim \mu_{\rm eff}\, T^2 \,.
\end{align}
As observed by Cohen and Kaplan, this result may serve as a basis for the successful generation
of the baryon asymmetry.
However, in order to arrive at a realistic model, one first of all
has to address three important questions:
(i) what is the nature of the field $a$ and the origin of the derivative coupling in Eq.~\eqref{eq:derivative},
(ii) how is the field $a$ set in motion, and
(iii) what kind of interactions drive the number density $n_F$
towards its equilibrium value $n_F^{\rm eq}$?
In the following, I shall discuss each of these issues in turn, cf.\ Sec.~\ref{sec:mechanism},
which will eventually lead us to an interesting alternative to thermal leptogenesis~\cite{Fukugita:1986hr}.
In Sec.~\ref{sec:parameters}, I will then sketch the parameter dependence
of the final baryon asymmetry in our model; and in Sec.~\ref{sec:conclusions},
I will conclude and give a brief outlook.

\section{Novel axion-driven leptogenesis mechanism}
\label{sec:mechanism}

\begin{figure}
\centering
\includegraphics[width=0.45\textwidth]{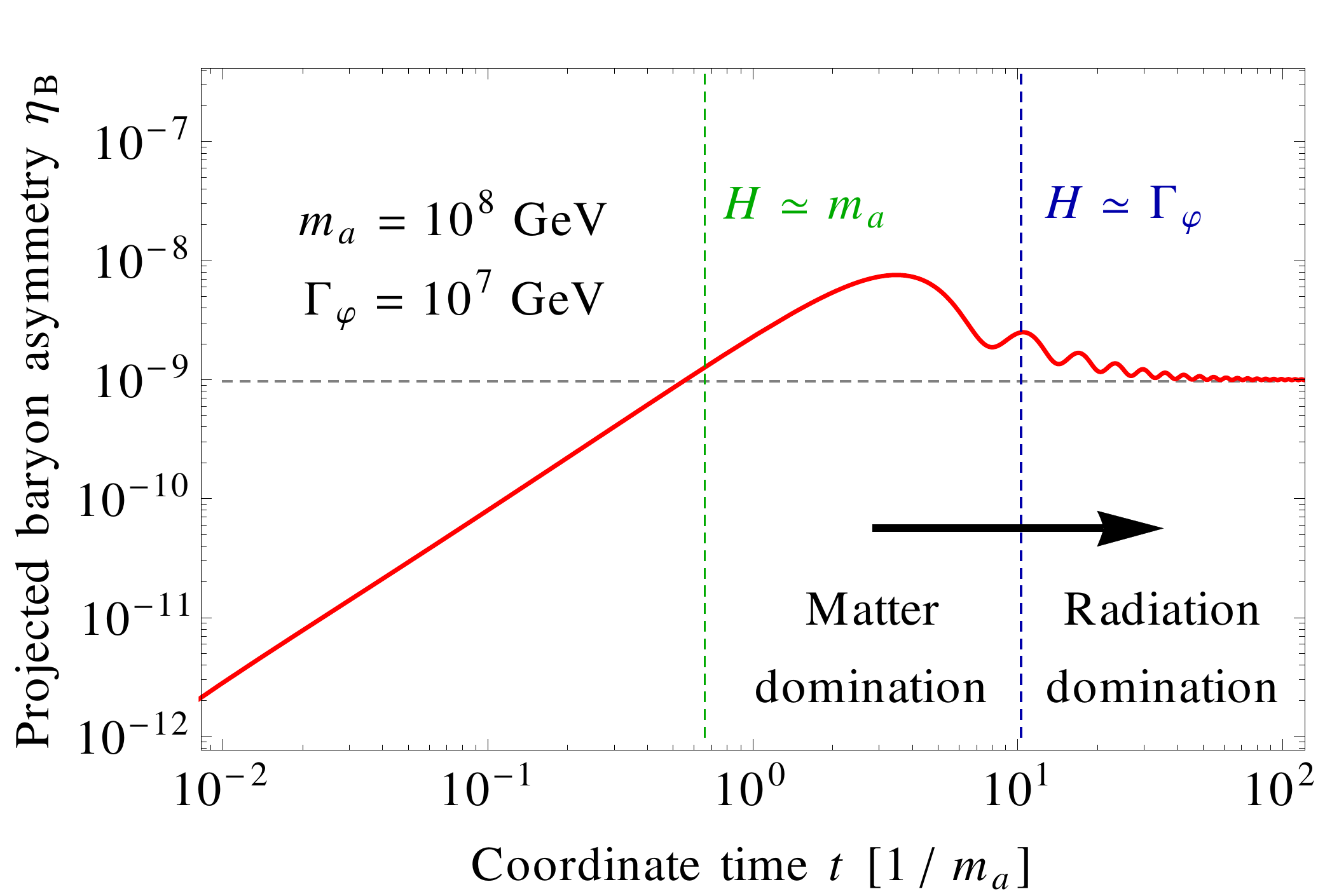}\hspace{0.05\textwidth}
\includegraphics[width=0.45\textwidth]{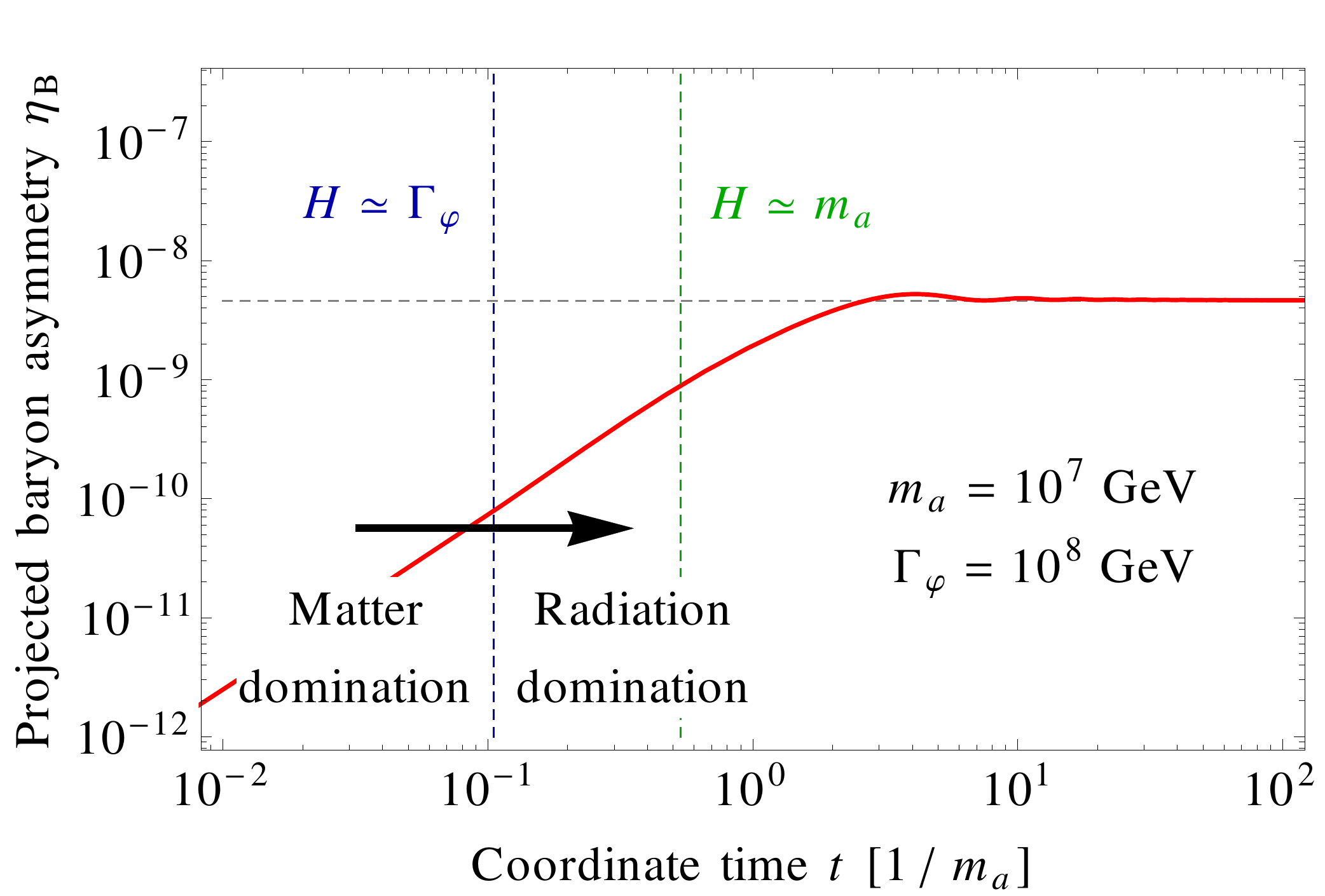}
\caption{Evolution of the (instantaneous) baryon asymmetry (projected onto its
would-be present-day value) as a function of time for
$m_a \gg \Gamma_\varphi$ \textbf{(left panel)} and
$m_a \ll \Gamma_\varphi$ \textbf{(right panel)}.
Here, $f_a = 3 \times 10^{14}\,\textrm{GeV}$ in both panels.
\label{fig:interplay}}
\end{figure}


Let us integrate the derivative interaction in Eq.~\eqref{eq:derivative} by parts,
$\mathcal{L} \supset 1/f_a\, \partial_\mu a \, j^\mu \rightarrow - a/f_a\, \partial_\mu j^\mu$.
This illustrates that the $CPT$-violating coupling required for baryogenesis
is equivalent to a coupling of the pseudoscalar $a$ to the divergence
of the fermion current $j^\mu$.
The field $a$ is thus naturally identified as an axion-like field, or simply ``axion''~\cite{Peccei:1977hh},
which couples to the anomaly of the fermion number $F = 3\,B + L$.
From this point of view, the cut-off scale $f_a$ in Eq.~\eqref{eq:derivative} is
immediately recognized as the decay constant of the axion field $a$.
Moreover, we note that the electroweak anomalies of the baryon number $U(1)_B$
and lepton number $U(1)_L$ in the SM allow us to recast the axion coupling to
$\partial_\mu j^\mu$ as a coupling to the electroweak field strength tensors
$W_{\mu\nu}$ and $B_{\mu\nu}$,
\begin{align}
\mathcal{L} \supset - \frac{a}{f_a}\,\partial_\mu j^\mu \rightarrow
\frac{a}{f_a} \frac{N_f}{8\pi^2} \left(g_2^2\, W_{\mu\nu}\tilde{W}^{\mu\nu}
- g_1^2\, B_{\mu\nu}\tilde{B}^{\mu\nu} \right) \,,
\label{eq:aFF}
\end{align}
where $N_f = 3$ is the number of SM fermion generations and with
$g_2$ and $g_1$ denoting the electroweak gauge couplings.
Interactions of this form may, for instance, arise in string theory,
which always features at least one (model-independent) axion~\cite{Witten:1984dg}
associated with the Green-Schwarz mechanism of anomaly cancellation~\cite{Green:1984sg}.
This axion couples to all gauge groups with universal strength
$f_a \sim 10^{16}\,\textrm{GeV}$~\cite{Choi:1985je}.
Besides that, string theory may also give rise to a multitude of further
axions fields coupling to different gauge groups with nonuniversal
strength~\cite{Witten:1984dg,Witten:1985fp}.
The couplings of these model-dependent axions are then determined by
the gauge structure as well as the details of the compactification.
For our purposes, the upshot of these considerations is that a certain linear
combination of stringy axions may very well end up coupling to $F\tilde{F}$,
where $F = W,B$.
In the following, we shall therefore identify the field $a$ in Eq.~\eqref{eq:derivative}
with just this linear combination and take the above string-based argument
to be the origin of the coupling $\mathcal{L} \supset a/f_a\,F\tilde{F} \leftrightarrow
1/f_a\, \partial_\mu a \, j^\mu \leftrightarrow \mu_{\rm eff} \, n_F$ in the Lagrangian.


The dynamics of the axion background in the early universe are governed by its
classical equation of motion,
\begin{align}
\ddot{a} + 3 \, H \, \dot{a} = - \partial_a V_{\rm eff}(a) \,, \quad
V_{\rm eff} \simeq \frac{1}{2} m_a^2 a^2 \,,
\label{eq:aEOM}
\end{align}
where the effective potential $V_{\rm eff}$ and hence the axion mass $m_a \simeq \Lambda_H^2 / f_a$
may, for instance, originate from instanton effects in a strongly coupled hidden
sector featuring a dynamical scale $\Lambda_H$.
Assuming that the PQ-like symmetry associated with the flat axion direction
is broken sufficiently early before the end of inflation (and not restored afterwards),
the initial axion field value $a_0 = (0\cdots 2\pi) f_a$ becomes constant on superhorizon
scales.
For definiteness, we shall therefore take $a_0 / f_a$ to be $1$ in the entire observable
universe at the end of inflation.
As we will see shortly, the baryon asymmetry produced during reheating is going to depend
on $a_0$.
Because of that, we have to ensure that the baryonic isocurvature perturbations induced by
the quantum fluctuations $\delta a$ of the axion field around its homogeneous background
$a_0$ remain below the observational bound~\cite{isocurvature}.
This constrains the Hubble rate $H_{\rm inf}$ during inflation:
$H_{\rm inf}/(2\pi)/a_0 \lesssim 10^{-5}$ or equivalently
$H_{\rm inf} \lesssim 6\times 10^{11}\,\textrm{GeV} \left(f_a / 10^{15}\,\textrm{GeV}\right)$.
At the same time, we have to require that $m_a \lesssim H_{\rm inf}$, so that
during inflation the Hubble friction term on the left-hand side of Eq.~\eqref{eq:aEOM}
outweighs the potential gradient on the right-hand side of Eq.~\eqref{eq:aEOM}.
After the end of inflation, the Hubble rate then begins to drop, until, around $H \simeq m_a$,
the axion begins to coherently oscillate around the minimum of its effective potential,
$a = 0$, with frequency $\omega = m_a$.
During this stage of \textit{axion relaxation}, the axion field therefore evolves with nonzero
velocity in its potential, which temporarily induces an effective chemical potential
for the fermion number, as anticipated at the beginning of this talk, cf.\ Eq.~\eqref{eq:mueff}.


\begin{figure}
\centering
\includegraphics[width=0.45\textwidth]{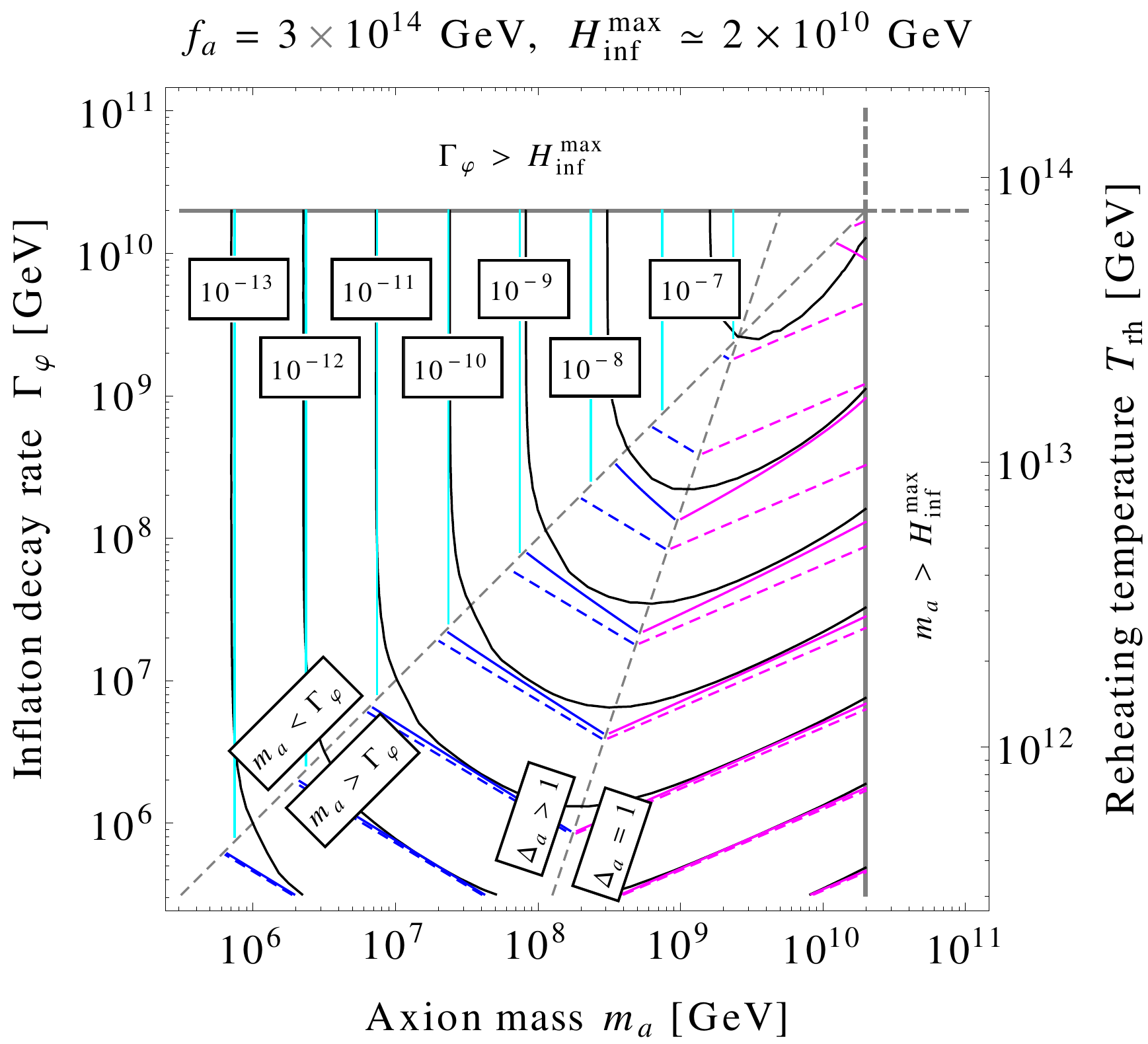}\hspace{0.05\textwidth}
\includegraphics[width=0.45\textwidth]{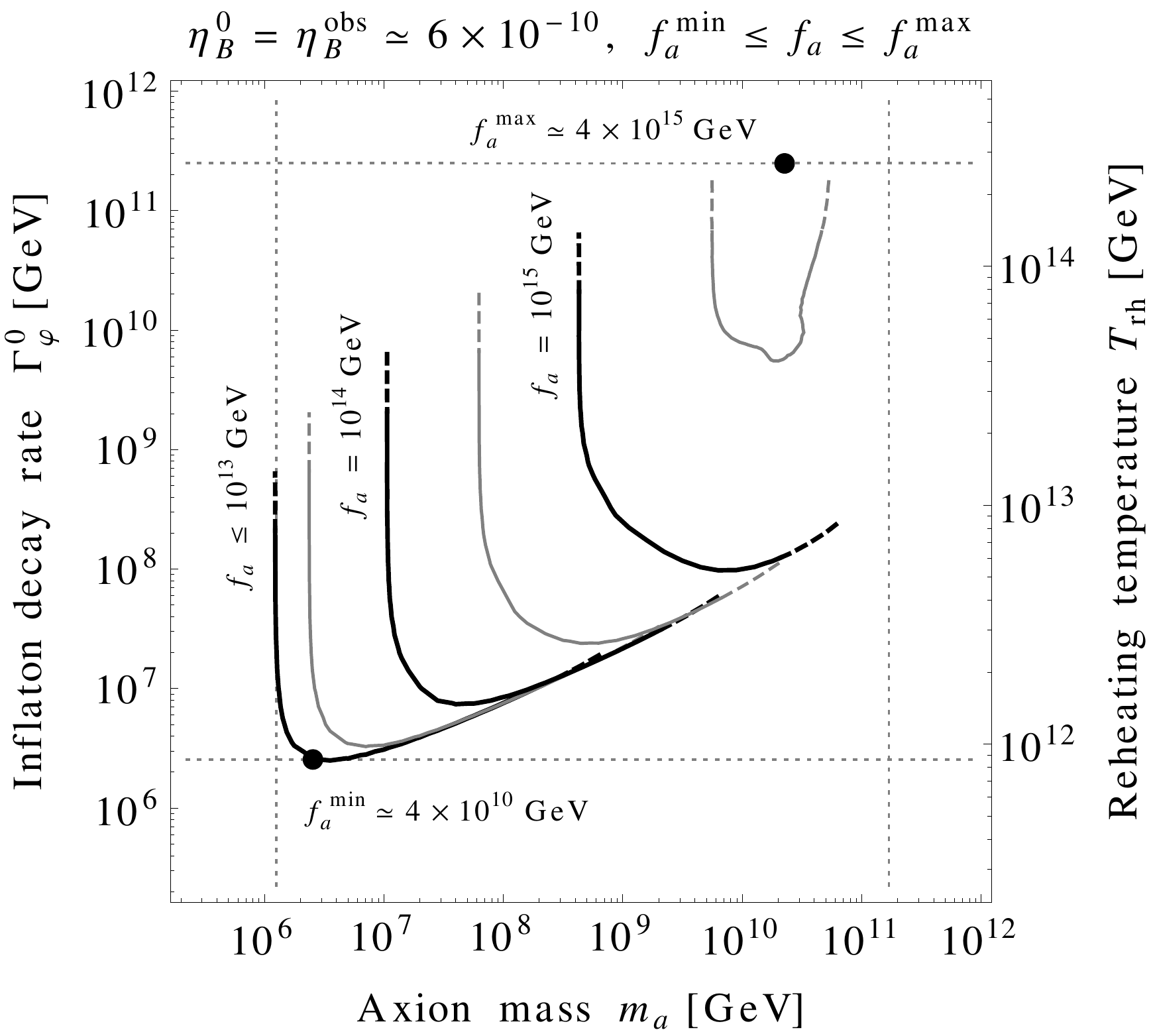}
\caption{\textbf{(Left panel)} Contour plot of the final baryon asymmetry $\eta_B^0$.
The black (bent) contours represent the full numerical result, while the colorful
(straight) contours depict the analytical estimate according to Eqs.~\eqref{eq:etaB},
\eqref{eq:etaL} and \eqref{eq:Delta}.
The effect of washout is illustrated by the difference between the dashed ($\kappa = 0$)
and solid ($\kappa \neq 0$) lines.
\textbf{(Right panel)} Contour lines of successful leptogenesis ($\eta_B^0 = \eta_B^{\rm obs}$)
for different values of $f_a$.
The dashed segments along the individual contours indicate where either $m_a$ or
$\Gamma_\varphi$ becomes comparable to the maximally allowed Hubble rate,
$H_{\rm inf}^{\rm max} \simeq 2\pi \,10^{-5}\,f_a$.
\label{fig:scan}}
\end{figure}


In order to make use of $\mu_{\rm eff} = \dot{a}/f_a$ for the purposes of baryogenesis,
it is important that there be an \textit{external source} of baryon or lepton number
violation that proceeds at a much faster rate $\Gamma$ than the motion of the axion field in its
effective potential, $\Gamma \gg \dot{a}/a$.
Only then do the axion oscillations act as an adiabatic background, so that $\dot{a}/f_a$
can be interpreted as an effective chemical potential~\cite{Dolgov:1994zq}.
Here, a minimal choice to satisfy this requirement is to rely on $L$ violation
through the $s$- and $t$-channel exchange of heavy Majorana neutrinos $N_i$,
\begin{align}
\Delta L = 2 : \quad \ell_i\ell_j \leftrightarrow N_k^* \leftrightarrow HH \,, \:\:
\ell_i H \leftrightarrow N_k^* \leftrightarrow \bar{\ell}_j\bar{H} \,, \quad 
\ell_i^T =
\begin{pmatrix}
\nu_i & e_i 
\end{pmatrix} \,,\:\: H^T =
\begin{pmatrix}
h_+ & h_0 
\end{pmatrix} \,, \:\: i,j,k = 1,2,3 \,.
\label{eq:scatterings}
\end{align}
These processes are guaranteed to be present in the bath as long as one
believes in the seesaw mechanism as the correct explanation for the small
neutrino masses in the SM~\cite{seesaw}.
In order to separate the leptogenesis mechanism under study form the contributions
from ordinary thermal leptogenesis, we shall assume that all right-handed neutrinos
$N_i$ acquire Majorana masses $M_i$ close to the scale of grand unification (GUT),
$M_i \sim \mathcal{O}\left(0.1\cdots 1\right)\Lambda_{\rm GUT} \sim 10^{15}\cdots 10^{16}\,
\textrm{GeV}$, so that none of them is actually ever thermally produced.
For center-of-mass energies $\sqrt{s}\ll M_i$, the thermally averaged cross section
of the $\Delta L = 2$ lepton-Higgs scattering processes in Eq.~\eqref{eq:scatterings},
$\sigma_{\rm eff} \equiv \left<\sigma_{\Delta L = 2}\, v\right>$, is then practically
fixed~\cite{Buchmuller:2004nz} by the experimental
data on the SM neutrino sector~\cite{Agashe:2014kda},
\begin{align}
\sigma_{\rm eff} \approx \frac{3}{32\pi} \frac{\bar{m}^2}{v_{\rm ew}^4}
\simeq 1\times10^{-31} \,\textrm{GeV}^{-2} \,, \quad \bar{m}^2 =
\sum_{i=1}^3 m_i^2 \approx \Delta m_{\rm atm}^2 \simeq 2.4 \times 10^{-3} \,\textrm{eV}^2
\,, \quad v_{\rm ew} \simeq 174 \,\textrm{GeV} \,.
\end{align}
Correspondingly, the evolution of the $L$ number density $n_L$ is described
by the following Boltzmann equation,
\begin{align}
\dot{n}_L + 3\,H\,n_L \simeq -\Gamma_L \left(n_L - n_L^{\rm eq}\right) \,, \quad
\Gamma_L = 4\, n_\ell^{\rm eq} \sigma_{\rm eff} \,, \quad
n_\ell^{\rm eq} = \frac{2}{\pi^2} T^3 \,, \quad
n_L^{\rm eq} = \frac{4}{\pi^2} \mu_{\rm eff} T^2 \,, 
\label{eq:Lboltz}
\end{align}
where we have approximated the lepton and $L$ number densities in thermal
equilibrium, $n_\ell^{\rm eq}$ and $n_L^{\rm eq}$, by their corresponding
expressions in the classical Boltzmann approximation.
Notice that the production term on the right-hand side of this equation,
$\Gamma_L n_L^{\rm eq} \propto \sigma_{\rm eff} \,\mu_{\rm eff}\,T^5$, is largely
independent of the details of the neutrino sector.
It does, in particular, not depend on the amount of $CP$ violation in the neutrino
sector nor on the heavy neutrino mass spectrum.
At the same time, it increases linearly with the light neutrino mass scale $\bar{m}$.
The usual bound on this mass scale from thermal leptogenesis, $\bar{m}\lesssim 0.2\,\textrm{eV}$,
(where it ensures that dangerous washout processes do not become too strong~\cite{Buchmuller:2005eh})
hence does not apply in our scenario.
The absolute neutrino mass scale will soon be probed experimentally
(on earth~\cite{Drexlin:2013lha} as well as in the sky~\cite{Abazajian:2011dt}).
This, therefore, entails the intriguing possibility
to test our model against the conventional scenario of thermal leptogenesis
in the near future.

\section{Parameter dependence of the final baryon asymmetry}
\label{sec:parameters}

Subsequent to its generation (according to Eq.~\eqref{eq:Lboltz}), the lepton asymmetry
is partly converted into a baryon asymmetry by means of electroweak sphalerons.
The final present-day baryon asymmetry $\eta_B^0$ is then given as
\begin{align}
\eta_B^0 = \frac{n_B^0}{n_\gamma^0} = c_{\rm sph} \, \frac{g_{*,s}^0}{g_*} \, \eta^L_a
\simeq 0.013 \, \eta_L^a \,, \quad
c_{\rm sph} = \frac{28}{79} \,, \quad g_{*,s}^0 = \frac{43}{11} \,, \quad 
g_* = \frac{427}{4} \,,
\label{eq:etaB}
\end{align}
where $c_{\rm sph}$, $g_*$ and $g_{*,s}^0$ denote the SM sphaleron conversion
factor as well as the effective number of relativistic degrees of freedom at high and
low temperatures, respectively.
Moreover, $\eta_L^a$ represents the final lepton asymmetry after the late-time decay of
the axion field at times around $t \sim \Gamma_a^{-1}$, where
$\Gamma_a \simeq g_2^2/(256\,\pi^4)\, m_a^3/f_a^2$.
In general, $\eta_L^a$ does not correspond to the equilibrium number density at the same
time, $\eta_L^{\rm eq}$, as the efficiency of leptogenesis typically begins to cease before
the equilibrium value is actually reached, so that $\eta_a^L \ll \eta_a^{\rm eq}$.


For fixed values of $a_0$ and $f_a$, the final lepton asymmetry ends up depending
on two parameters:
the axion mass $m_a$ as well as the reheating temperature $T_{\rm rh}$, which is, in turn,
determined by the inflaton decay rate,
$T_{\rm rh} \simeq 0.3\,\sqrt{\Gamma_\varphi M_{\rm Pl}}$.
We infer the precise parameter dependence of $\eta_L^a$ by numerically
solving Eqs.~\eqref{eq:aEOM} and \eqref{eq:Lboltz} together with the
Friedmann equation for the scale factor as well as the Boltzmann equations
for the number densities of inflaton particles and relativistic SM particles, respectively.
As it turns out, our exact numerical result can be very well reproduced by
the following analytical expression, cf.\ also the left panel of Fig.~\ref{fig:scan},
\begin{align}
\eta_L^a = C \, \Delta_a^{-1} \Delta_\varphi^{-1}\,\eta_L^{\rm max}\, e^{-\kappa} \,, \quad
\eta_L^{\rm max} \simeq \frac{\sigma_{\rm eff}}{g_*^{1/2}} \frac{a_0}{f_a}
\, m_a \, M_{\rm Pl}
\times \min\left\{1,\,\left(\Gamma_\varphi/m_a\right)^{1/2}\right\} \,,
\label{eq:etaL}
\end{align}
with $C$ being a numerical fudge factor of $\mathcal{O}(1)$.
$\eta_L^{\rm max}$ denotes the all-time maximum of the lepton asymmetry, which
is reached around the time when the axion oscillations set it, i.e., at
$t \sim t_{\rm osc} \simeq m_a^{-1}$.
Note that it is rather insensitive to both $a_0$ and $f_a$, as it only
depends on the ratio $a_0/f_a$, which is expected to be of $\mathcal{O}(1)$.
Furthermore, $\Delta_\varphi$ and $\Delta_a$ account for the dilution of $\eta_L^{\rm max}$
in the course of inflaton and axion decays, respectively,
\begin{align}
\Delta_\varphi \simeq \max\left\{1,\,\left(m_a/\Gamma_\varphi\right)^{5/4}\right\} \,, \quad
\Delta_a \simeq \max\left\{1,\,\frac{8\pi^3}{g_2^2}\frac{f_a\, a_0^2}{m_a\, M_{\rm Pl}^2}
\times \min\left\{1,\,\left(\Gamma_\varphi/m_a\right)^{1/2}\right\}\right\} \,.
\label{eq:Delta}
\end{align}
Here, $\Delta_\varphi$ reflects the interplay between leptogenesis and reheating, cf.\
Fig.~\ref{fig:interplay}.
For $m_a \gtrsim \Gamma_\varphi$, the axion begins to oscillate before the end of reheating
and the initial asymmetry becomes diluted due to the entropy production in inflaton decays. 
For $m_a \lesssim \Gamma_\varphi$, on the other hand, the axion oscillations only set in after
the end of reheating and the final asymmetry becomes independent of the inflaton decay rate.
Meanwhile, $\Delta_a$ begins to play a role for $f_a$ values around
$3\times10^{13}\,\textrm{GeV}$, cf.\ the right panel of Fig.~\ref{fig:scan}.
For smaller values of $f_a$, we always have $\Delta_a = 1$ in the entire parameter region
of interest.
The factor $e^{-\kappa}$, finally, accounts for the efficiency of the washout term,
$-\Gamma_Ln_L$, on the right-hand side of Eq.~\eqref{eq:Lboltz}.
For $m_a \gtrsim \Gamma_\varphi$, $\kappa$ can be roughly estimated as
$\kappa \sim T_{\rm rh}/T_L$, where
$T_L \sim 1/\left(\sigma_{\rm eff} M_{\rm Pl}\right) \sim 10^{13}\,\textrm{GeV}$
is the typical temperature scale of leptogenesis, while, for $m_a \lesssim \Gamma_\varphi$,
we have $\kappa \sim 1$.
A better analytical understanding of washout in our scenario is, however, still pending.


Successful leptogenesis restricts the axion decay constant $f_a$ 
to take a value within the following range,
\begin{align}
4\times 10^{10}\,\textrm{GeV} \lesssim f_a \lesssim 4\times 10^{15}\,\textrm{GeV} \,.
\end{align}
which translates into allowed ranges for $m_a$, $\Gamma_\varphi$ and $T_{\rm rh}$,
cf.\ the right panel of Fig.~\ref{fig:scan}, which are all very well consistent with
typical string axion models.
We note that, for smaller values of $f_a$, it is not possible to generate a sufficiently large
baryon asymmetry, while keeping the baryonic isocurvature perturbations small enough.
Likewise, for larger values of $f_a$, the dilution of the asymmetry in the late-time decay of
the axion is too strong.

\section{Conclusions and outlook}
\label{sec:conclusions}

While thermal leptogenesis typically operates at
$T_{\rm rh} \sim 10^{10}\,\textrm{GeV}$, the requirement of a large rate
of $L$ violation, $\Gamma_L \gg H$, pushes $T_{\rm rh}$
to values at least of $\mathcal{O}\left(10^{12}\right)\,\textrm{GeV}$ in our scenario.
Furthermore, our final baryon asymmetry turns out be independent of the amount of $CP$
violation in the neutrino sector as well as of the $N_i$ mass spectrum.
On top of that, the usual bound on
$\bar{m}$ from thermal leptogenesis does not apply in our case.
The presented model should therefore be regarded as
an attractive alternative to thermal leptogenesis in case the latter should begin to look less
favorable from the experimental point of view!
Beyond that, further work is needed:
it remains, for instance, to be seen how the required high $T_{\rm rh}$
could be possibly accommodated in a supersymmetric version of our model.
A further intriguing question, which we are currently investigating, is whether
the role of the axion field $a$ could not be equally played by the inflaton.
This would result in an even more minimal scenario.


\begin{acknowledgments}
I wish to thank A.~Kusenko and T.~T.~Yanagida for many helpful discussions and the
fruitful collaboration on our joint paper; I wish to thank the organizers of
\textit{HPNP 2015} for a wonderfully organized workshop; and I wish to thank 
all participants of \textit{HPNP 2015} for an exciting week in Toyama.
In particular, I am grateful to T.~T.~Yanagida for generous travel support
through Grants-in-Aid for Scientific Research from the Ministry of Education,
Science, Sports, and Culture (MEXT), Japan, \#26104009 and \#26287039.
In addition, this work has been supported in part by the World Premier International Research
Center Iniative (WPI), MEXT, Japan.
\end{acknowledgments}

\bigskip 

\end{document}